# Rigel Exoplanet Geologist

A white paper submitted to
Solar and Space Physics (Heliophysics) Decadal Survey (2022) of
The National Academies of Sciences

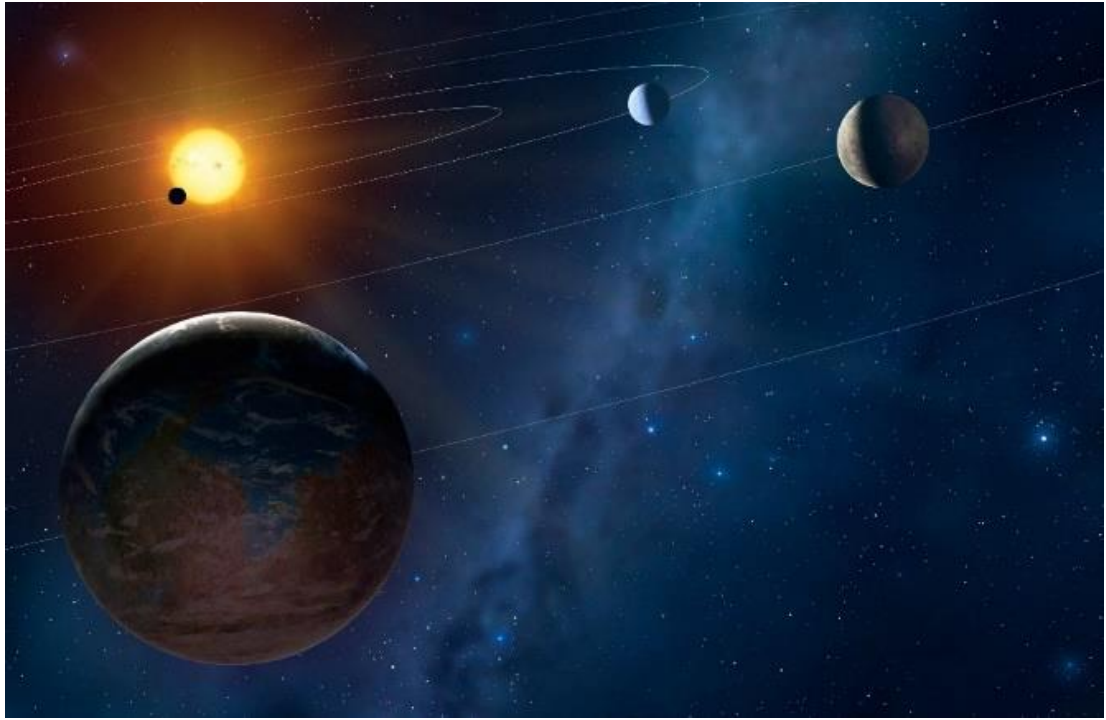

Philip Horzempa
LeMoyne College; Syracuse, New York
September 7, 2022

Email: horzempa45@gmail.com

**RIGEL EXOPLANET GEOLOGIST**
September 7, 2021
Philip Horzempa

**Synopsis**

   The next great objective in Space is the surface exploration of a planet orbiting a distant star. The Robotic Interstellar GEologicaL probe, RIGEL, is designed to land on an exoplanet and roam across its surface.  This will be a logical follow-on to space telescope surveys and flyby missions.  The engineering challenge that this presents is one reason that the RIGEL project should be initiated promptly.  The hurdles are immense, but the prize is worth the effort.  For the first time in history, an explorer from the Earth will be able to walk about the surface of an alien planet.  That explorer will be a machine, a proxy for mankind.
   Besides geology, Rigel will also analyze any life on its target world.  There are investigations that can only be conducted in-situ.  For example, if there are plants which use stellar light (sunshine) as an energy source, the structure of their photosynthesizing pigment can only be determined by directly analyzing the plant structure.

   This is a mission that will span centuries.  It does not assume undefined breakthroughs that would allow rapid transit across interstellar space.  The Rigel concept exists within the known laws of physics.  We can send a robot proxy to a distant exoplanet, but patience is required.  The daunting chasm between stars dictates the major features of the Rigel craft.  It must be built to last for centuries and it must act completely on its own.
   The Decadal Survey is asked to recommend the creation of a new Program within NASA to manage this effort over the next 1,000 years.  Once Project Rigel has commenced, a tangible program will exist, one which other missions can contribute to.  The Mars Exploration Program could cite their efforts to develop increasingly capable rovers.  One of the science drivers for future Space telescopes will be their role in determining which exoplanet should be Rigel's target.
   Technology will improve over the next 200 years, but that is precisely what the Rigel project is positioned to utilize.  Plans may change, but you can't change a plan unless you have a plan.  NASA should begin pre-formulation (pre-Phase A) efforts on the Rigel probe by 2029.
   As with the Apollo Moon landing project, details regarding hardware and mission design will evolve.  In a similar fashion, the end goal for Rigel will be simple but profound, i.e., landing a geologist avatar on an exoplanet.

   One school of thought posits that it is best to wait to begin our trek to the stars because undefined technology of the future will allow the construction of faster and faster vehicles.  Why bother launching a probe with existing technology when it will be overtaken by spacecraft utilizing fictional technology?  This paper does not hold with that philosophy.  There is no sign that undiscovered laws of physics are waiting to be discovered.  The limits that are now apparent are, more than likely, permanent limits.  The best course of action is to summon our resources and go about the job of taking the first step towards fashioning an interstellar explorer.

Some hardware elements will be amenable to near-term development.  Others will require development over decades.  The existence of an official project office will allow those efforts to proceed in a more organized and directed manner.

Once set in motion, this project will last for hundreds of years.  That time span can be daunting, but success can be achieved with patience.  However, some generation must begin that process.  The Decadal Survey team should recommend that this generation take that first step.  There is nothing to be gained by waiting.

**Concept Outline**

In order to understand a world, one needs a geologist.  For alien planets, that geologist must be a robot.  Rigel will seek to uncover the Deep Past of its target planet.  As on the Earth, this can only be done by the direct examination of the Rock Record.   This proposal will use a planet in the tau Ceti system as its target.  At a distance of 10 light-years, this may be the nearest system that includes a temperate rocky planet.

The first challenge is propulsion.  It will be assumed that the Rigel probe can be accelerated to a velocity of 2,000 miles per second (3,200 km/sec).  That is slightly greater than 1% of the speed of light (c).  The Pluto Express (New Horizons) spacecraft, with a mass of 2,000 pounds (1,000 kg), achieved a velocity of 10 miles per second (16 km/sec).

Let's assume that Rigel will also have a mass of 2,000 pounds, but with a velocity that is 200 times greater.   That will result in a kinetic energy level for Rigel that is 40,000 times as great as that for New Horizons.  Achieving that level of energy will require a focused engineering effort.

Another challenge will be communications over a distance of 60 million million miles (10 light-years).  That is 1 million times the average distance to Mars.  Since radiated energy levels drop with the square of distance, Laser systems may be the best way to achieve that interstellar data link.  Time will be a big factor.  At 1% of the speed of light, the Rigel probe will take 1,000 years to reach the tau Ceti system.

**Implementation**

NASA's planning horizon varies from 5 to 20 years.  For RIGEL to become a reality, that horizon must encompass the next 100 to 1,000 years.  Without that adjustment, such a mission will never earn a New Start.  This change is required because a robot such as Rigel will require technology development will span decades or even centuries.

All of this will be preceded by global mapping of the target planet from the confines of the Solar System.  Space telescopes that can produce a 1st-order map might be of several varieties.  One would be the Exo-Earth Mapper that uses an array of telescopes with the technique of Interferometry to produce images.  Another effort would send an array of small telescopes to the Solar Gravity Lens at 500 AU.

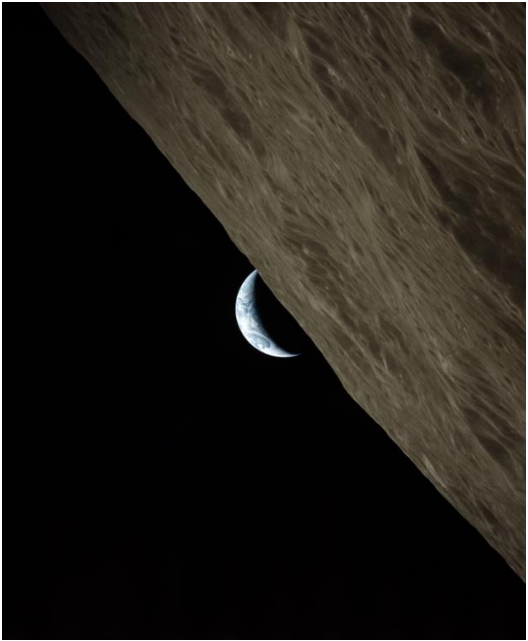
Rounding an Exo-Moon on Approach

More challenges include:
> spacecraft that can hibernate for 100 years
> power systems that can function for 1,000 years
> avionics that can last for 1,000 years
> Whipple shields that can withstand the impact of particles moving at 2,000 miles per second

There are already several efforts that will contribute to the Rigel concept.  The MSL and Mars 2020 rovers have advanced the science of proxy geologists on another planet.  They have demonstrated low-mass instruments, improved mobility and increasing levels of independent activities.

To enable the Rigel explorer, a series of precursors will be flown within the solar system.  To rehearse the concept of complete autonomy, rovers will explore Baja California for two years with no outside intervention.  Humans can observe, and record, the progress of the vehicles but will not interfere.  The next step will be sending rovers to Mars with the craft operating independently from the moment of separation from the launch vehicle.  The craft will use aerocapture to enter orbit around Mars.  It will then survey the planet and independently choose a landing site.  It will then leave orbit, conduct a high-speed entry, and land on Mars on its own volition, choosing when and where it will land.

After touchdown, the rover will roam over 100s of miles of Martian terrain in the span of two years.  This will all be done without input from the Earth.  The rover will explore a location, then move on.  Its next target will be chosen by the onboard computer using data from the ground as well as images taken from orbit.  Its goal will be to explore as much of the surface as it can every day, every month and every year.  This dry run on Mars will set the stage for programming the Rigel spaceship.

**Propulsion**

There are several candidates for propelling the Rigel probe to the tau Ceti system. Perhaps the most likely method of acceleration harkens back to the Orion nuclear spaceship proposed in the early years of the Space Age.  That would have been a massive vehicle, whereas a robot such as Rigel will be a comparative lightweight.  Orion's use of pulsed thermonuclear explosion shockwaves as a means of propulsion may be the only pathway for Rigel to achieve a velocity of 0.01c.  This is where an approved project would allow studies into a more streamlined approach.  Research will be conducted into making the nuclear devices as small as possible, and minimizing the mass of a pusher shield.

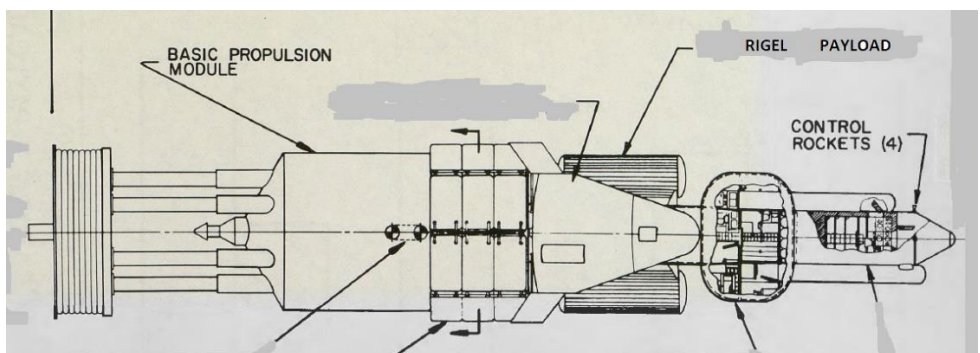
Nuclear Pulse Spacecraft (credit: General Dynamics)

The problem of deceleration into orbit around tau Ceti is the other end of the propulsion challenge.  It may be that the use of the star's magnetic field will allow a capture orbit, followed by several years of "mag-breaking" to enter an orbit near the target planet.
The above discussion of propulsion options illustrates the challenges facing the Rigel project.  They could be viewed as insurmountable, reasons to not start the project.  However, the opposite is true.  With an approved project and an annual budget, a near-term start will provide the best chance to solve those engineering puzzles.

**Mission Details**

Since no samples will be returned to Earth, Rigel will carry a miniature geology laboratory. Key features to be chronicled include stratigraphic sequences and absolute ages of those rock layers.  As on the Earth, exposures such as escarpments will allow Rigel to map rock outcrops.
This will include an automatic thin-section device that will be crucial for the operation of the other instruments. An in-situ geochronology lab is vital to uncovering the Deep Past of the target planet.

The mission sequence is as follows:
> departure from the Solar System at 2,000 miles/sec
> hibernation mode for 1,000 years
> contact with Earth every 20 years
> awakening from hibernation 1 year before arrival in tau Ceti system
> solar panels deployed 1 week before at tau Ceti
> deceleration into orbit around tau Ceti
> entry and landing on target planet in tau Ceti system
> deployment of mobile robot geologist
> exploration of planet's geology and any life

  During the phase of approaching the target planet's orbit, the probe will chart the magnetic field, stellar wind and other aspects. These in-situ measurements will allow a detailed comparison with our Sun. Essentially, this will be an interstellar version of a Heliophysics mission.
  To prepare the ground, so to speak, a series of **Sprinter** probes will be sent to the target system. These will be miniature spacecraft, on the order of 20 to 100 kg. They will perform fast fly throughs of the planetary system of interest, moving at 4,000 miles per second. Their task will be to construct a detailed map of the target exoplanet. A series of these Sprinters will be launched in the 100 years preceding the launch of Rigel. They will serve as redundant probes in case of failures of other craft in this class.

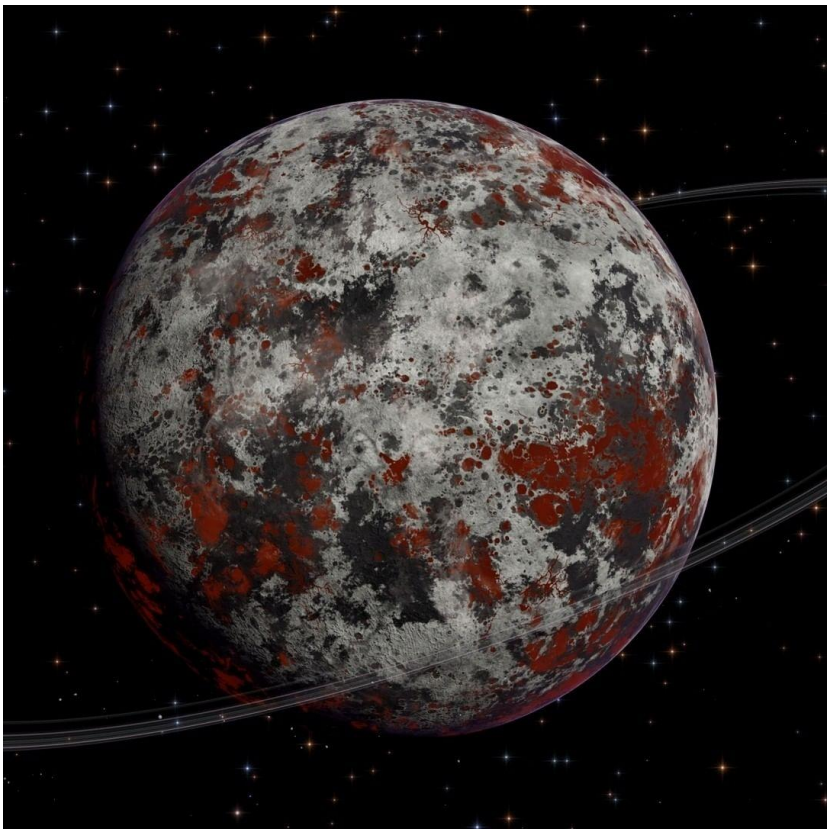

The Decadal Survey should direct NASA to increase the funding and scope of the Exoplanet Exploration program.   This was a field of study that did not exist 30 years ago.  The astronomy and geology communities are still in the process of adjusting to its rapid growth.  The first step would be the production of a report detailing what needs to be done over the next 200 years.

   This will require NASA to adjust how it plans for future Science missions.  Not only does the Rigel mission extend past the planning horizon for the federal budget.  It also extends beyond the next several election cycles.  That calls for very long-term planning, on the order of centuries.  This significant change in planning philosophy is why it is important that the Decadal Survey recommends this new approach to NASA.

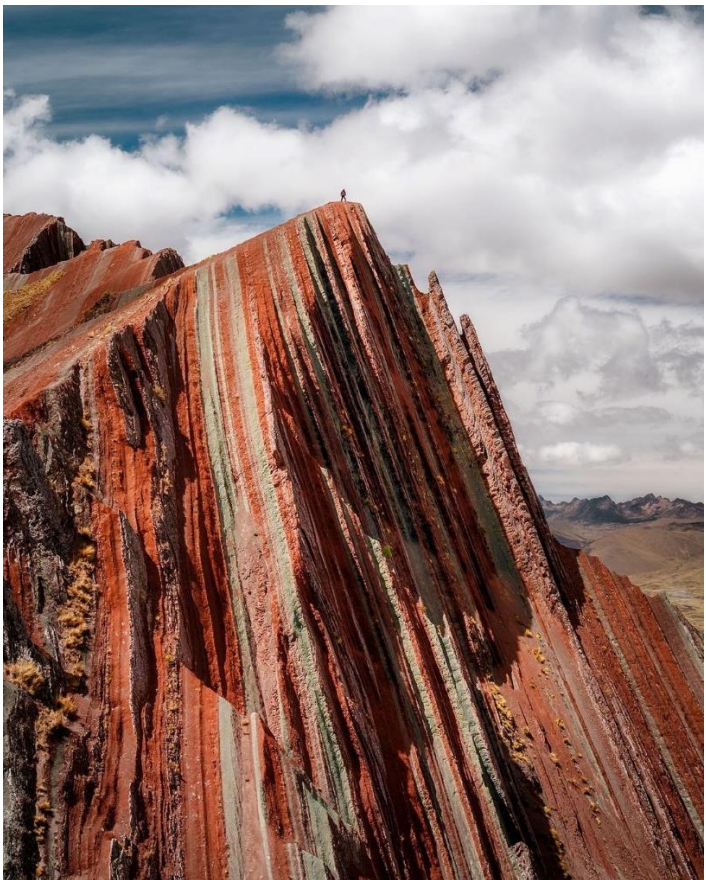
Rigel Proxy on Exoplanet Rock Strata

**Conclusions**

   As with Apollo, the Rigel project will provide an umbrella that will marshal the efforts of many fields of engineering.  They will be inspired because they will know that they are contributing to a noble, long-term goal of all of mankind.  To harken back to the words of JFK when referring to an earlier challenge, Rigel will "serve to organize and measure the best of our energies and skills."  Apollo accomplished its goal within a decade.  Rigel will require centuries.  That is the task that lies before us if we, via our robot proxies, are ever to explore alien planets.  There is no magic solution to the puzzle that is interstellar travel.  The stars and their planets lie across an almost unfathomable gulf of space.  That is a barrier that cannot be changed.  The job of this generation is to the realize the scale of the task, then go about the business of getting started.  This is the legacy that we can leave to future generations.  At some point, the torch will be passed to them, as they will later pass it to those that follow.

**References**

Nuclear Pulse Propulsion; 1965; Hill; General Dynamics

Simulation study of solar wind push on a charged wire: basis of solar wind electric sail propulsion; Janhunen, P. and Sandroos, A.; Annales Geophysicae 2002; Volume 25, pp. 755-756